\def\mytitle{Diffusing-Wave Spectroscopy in a Standard Dynamic Light Scattering Setup}
\def\authorone{Zahra Fahimi}

\def\authortwo{Frank Aangenendt}

\def\authorthree{Panayiotis Voudouris}

\def\authorfour{Johan Mattsson}

\def\lastauthor{Hans M. Wyss}

\def\tuemecheng{Department of Mechanical Engineering, Materials Technology, Eindhoven University of Technology, Eindhoven, the Netherlands}
\def\tueicms{Institute for Complex Molecular Systems, Materials Technology, Eindhoven University of Technology, Eindhoven, the Netherlands}
\def\tuedpi{Dutch Polymer Institute (DPI), Eindhoven, the Netherlands}
\def\leedsphys{School of Physics and Astronomy, University of Leeds, Leeds, United Kingdom}
\def\myabstract{Diffusing-Wave Spectroscopy (DWS) extends dynamic light scattering measurements to samples with strong multiple scattering. DWS treats the transport of photons through turbid samples as a diffusion process, thereby making it possible to extract the dynamics of scatterers from measured correlation functions. The analysis of DWS data requires knowledge of the path length distribution of  photons traveling through the sample. While for flat sample cells this path length distribution can be readily calculated and expressed in analytical form, no such expression is available for cylindrical sample cells.  DWS measurements have therefore typically relied on dedicated setups that use flat sample cells. 
Here we show how DWS measurements, in particular DWS-based microrheology measurements, can be performed in standard dynamic light scattering setups that use cylindrical sample cells.  To do so we perform simple random walk simulations which yield numerical predictions of the path length distribution as a function of both the transport mean free path and the detection angle. This information is used in experiments to extract the mean-square displacement of tracer particles in the material, as well as the corresponding frequency-dependent viscoelastic response. An important advantage of our approach is that by performing measurements at different detection angles, the average path length through the sample can be varied.  Using measurements on a single sample cell, this gives access to a wider range of length and time scales than obtained in a conventional DWS setup. Such angle-dependent measurements also offer an important consistency check, as for all detection angles the DWS analysis should yield the same tracer dynamics, even though the respective path length distributions are very different. We validate our approach by performing measurements both on aqueous suspensions of tracer particles and on solid-like gelatin samples, for which we find our DWS-based microrheology data to be in very good agreement with rheological measurements performed on the same samples.}
\def\mypacs{47.57.-s, 46.35.+z, 78.35.+c, 07.60.-j}

\def\MSD{\left<\Delta r^2(t)\right>}  
\def\lstar{{l^{\star}}}

\documentclass[pre,twocolumn,showpacs,superscriptaddress,floatfix,nofootinbib]{revtex4}

\usepackage{standalone}
\usepackage{graphicx}
\usepackage{dcolumn}
\usepackage{bm}
\usepackage{listings} 

\usepackage{color}

\begin{document}

\preprint{preprint number}

\title{\mytitle}

\author{\authorone}
\affiliation{\tuemecheng}
\affiliation{\tueicms}

\author{\authortwo}%
\affiliation{\tuemecheng}
\affiliation{\tueicms}
\affiliation{\tuedpi}

\author{\authorthree}
\affiliation{\tuemecheng}
\affiliation{\tueicms}

\author{\authorfour}
\affiliation{\leedsphys}

\author{\lastauthor}
\email{H.M.Wyss@tue.nl}
\homepage{http://www.mate.tue.nl/~wyss}
\affiliation{\tuemecheng}
\affiliation{\tueicms}
\affiliation{\tuedpi}

\date{\today}

\begin{abstract}
\myabstract
\end{abstract}

\pacs{\mypacs}
\maketitle

\section{Introduction}
\label{introduction}

Since its development in the 1980's, Diffusing-Wave Spectroscopy (DWS)~\citep{Maret:1987wl,Weitz:vp,Pine:1990bj} has proven to be an important and versatile tool for studying the dynamics, mechanics and structure of a wide range of soft materials.~\citep{tenGrotenhuis:2000js,Harden:2001ds,CohenAddad:2001jh,nr641,Weitz:1989tf,Weitz:1993tc,Scheffold:2002jb,Wyss:2001wj}
By taking advantage of the fact that the transport of photons through an optically turbid sample can be described as a diffusion process, DWS extends Dynamic Light Scattering (DLS) measurements to the highly multiple scattering regime. It thus enables access to the dynamics of a material at very short time and length scales.
The method is particularly useful when combined with the concept of \emph{microrheology}, where information on the dynamics of tracer particles added to a material are used to extract information on the material's viscoelastic properties.~\citep{Mason:1995jh,Mason:1997uz,Mason:1996hm,Mason:2000wz,Dasgupta:PhysRevE:2002} However, the proper analysis of any DWS measurement requires detailed knowledge of the path length distribution $P(s)$ for photons traveling through the sample to the detector.
For a number of sample geometries and experimental situations, the calculation or estimation of $P(s)$ has been described in previous studies, including for the situation of backscattering from a flat sample cell of infinite thickness, or for transmission through cone-plate cells or flat circular cells of finite diameter and thickness.~\citep{Kaplan:1993cj,nr441,Weitz:1993tc} Importantly, for sample cells in the shape of a flat slab of thickness $L$, infinitely extended in height and width, $P(s)$ can be expressed in analytical form, and the analysis of DWS data is therefore straightforward.~\citep{Pine:1990bj,Kaplan:1993cj}\\
For the cylindrical sample cells used in conventional dynamic light scattering setups, however, an analytical expression for $P(s)$ is not available. DWS measurements are therefore usually performed in dedicated instruments that use flat sample cells. 

In this paper, we show how DWS-measurements can be performed in a standard dynamic light scattering setup, using cylindrical sample cells. We perform simple numerical random walk simulations to account for the propagation of photons through a cylindrical cell and describe how this information is used to obtain the mean-square displacement of tracer particles from the temporal autocorrelation functions determined in experiments. We further show that by performing measurements at different detection angles, the range of accessible time and length scales can be extended; this is in analogy to standard DWS measurements employing a range of different cell thicknesses.
Importantly, our approach also provides a valuable consistency check, especially in the context of microrheology, since measurements taken at different detection angles should yield the same viscoelastic response, even though the corresponding correlation functions must be very different due to the variation in geometry and average path length.

In analogy to a conventional DWS measurement, the transport mean free path $\lstar $ is determined by a calibration measurement on tracer particles of well known uniform size, suspended in a Newtonian liquid of known viscosity; the expected single particle dynamics is thus known \emph{a priori}. Alternatively, the transport mean free path can also be directly determined from the measured scattering intensity as a function of angle, $I(\theta)$ if an initial experimental calibration is combined with results from our simple numerical calculations. In the highly multiple scattering limit, and in the absence of absorption in the sample, $I(\theta)$ is well approximated by a function that depends only on $\lstar $, on the corresponding calculated path length distribution $P(s)$, and on a constant $\beta_{\mathrm{exp}}$ that is determined by the experimental setup.

We thus demonstrate that standard goniometer light scattering setups can be successfully used for DWS measurements. Compared to dedicated DWS setups, our method has the advantage of being able to reliably determine the transport mean free path $\lstar $ as well as to extend the range of accessible length and time scales, using only a single cylindrical sample cell. 

We illustrate and test the use of our approach by performing DWS-based microrheology measurements on a typical solid-like soft material, gelatin, and find the resulting frequency-dependent viscoelastic moduli to be in very good agreement with separately performed rheological measurements.

\section{Experimental: Materials and Methods}
\label{experimental:materialsandmethods}

\begin{figure}[t] \centering
\includegraphics[width=\linewidth]{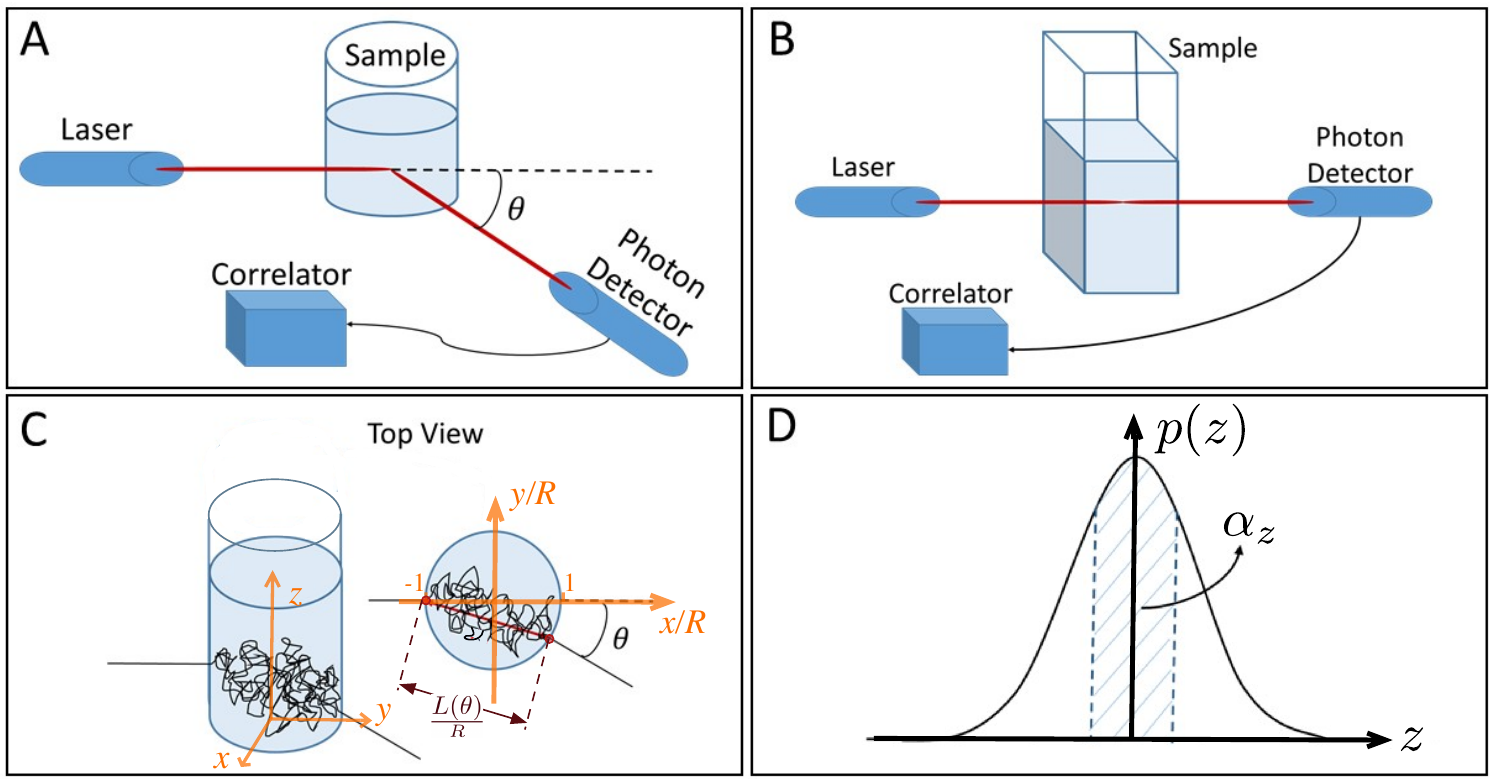}
\caption{(Color online) Schematic experimental setup for typical dynamic light scattering measurements. \textbf{(A)} \emph{Standard dynamic light scattering} (DLS) setup employing a cylindrical sample cell and a goniometer, which enables accessing different scattering angles $\theta$. \textbf{(B)} \emph{Diffusing-Wave Spectroscopy} (DWS) setup in transmission geometry. The pathways of photons are well-described by a random walk of step size $\lstar$. \textbf{(C)} Schematic of random walk simulations in a cylindrical geometry. The detection angle $\theta$ is defined as for conventional DLS; here it determines the distance $L(\theta)$ between the points of entry and exit of detected photons.  \emph{(D)} Expected distribution $p(z)$ of the $z$ coordinate where a photon exits the cylinder; a fraction $\alpha_z$ (marked area) reaches the detector.}
\label{Fig1}
\end{figure}

\subsection{Background on DLS, DWS, and microrheology}
\label{backgroundondlsdwsandmicrorheology}

Standard static and dynamic light scattering experiments are limited to samples that exhibit very little multiple scattering, with the overwhelming majority of detected photons having been scattered only a single time within the sample. A typical setup for such single scattering experiments uses a cylindrical sample cell that is illuminated by a laser, as shown schematically in Fig.\ref{Fig1}(A). The detector, typically comprising an optical fiber that is coupled to a photomultiplier tube, can be positioned at a range of detection angles $\theta$, corresponding to scattering wave vectors $q(\theta)=\frac{4 \pi n}{\lambda}   \sin(\theta/2)$, where $n$ is the refractive index of the sample and $\lambda$ is the wavelength of the laser in vacuum.
For single scattering, the fluctuations in the detected intensity, which reflect the dynamics of the scatterers, are then quantified by the temporal intensity autocorrelation function
\begin{equation}
\label{g2def}
g_2(t)=\frac{{< I(\tilde{t}+t)  I(\tilde{t}) >_{\tilde{t}} } }{ {< I(\tilde{t}) >_{\tilde{t}}}^2 } \ ,
\end{equation}
where $t$ is the lag time and the brackets $< .. >_{\tilde{t}}$ indicates a time-average over all times $\tilde{t}$.
The field autocorrelation function $g_1(t)$, measured at a wave vector $q$, reflects the temporal fluctuations of the electric field. It can be related, via the so-called Siegert relation to the intensity correlation function, as $g_1(t) \approx \sqrt{\left(g_2(t) - 1\right)/\beta}$, where $\beta$ is the coherence factor,~\citep{Berne:wf} a constant that depends on the experimental setup.
For a Gaussian distribution of displacements $\Delta r$, the field correlation function $g_1(t)$ is directly linked to the dynamics of scatterers in the sample, as 

\begin{equation}
\label{g1_t_single scattering}
g_{1}(t)=e^{-\frac{{q}^2}{6} \MSD} ,
\end{equation}

where $\MSD$ is the time-dependent mean-square displacement of scatterers in the material. For the simplest example, where the scatterers are uniformly sized particles suspended in a Newtonian liquid, the particles undergo ideal Brownian motion, and thus $\MSD = 6 D  t$, where $D$ is the particle diffusion coefficient in 3 dimensions. For this case, the field correlation function has a single exponential form, $g_1(t) = e^{- \Gamma t}$, where the $q$-dependent decay rate $\Gamma = D q^2$ is set by $D$.

\emph{Diffusing-Wave Spectroscopy} (DWS) is an extension of dynamic light scattering measurements to the highly multiple scattering regime. A typical experimental setup for DWS is shown in Fig.\ref{Fig1}(B). In contrast to conventional DLS measurements, this technique \emph{requires} that photons are scattered many times, before they reach the detector. For this highly multiple scattering regime, the propagation of photons through the sample can be adequately described as a simple diffusion process, where the details of each single scattering event are no longer relevant. This photon diffusion processes can be accounted for by a single parameter, the so-called transport mean free path $\lstar$. This characteristic length scale is defined as the average distance a photon travels in the sample before its direction of propagation is randomized. The path of photons through the sample can thus be approximated as an ideal random walk with step size $\lstar$. For such a random walk the path length of photons, and the number of randomizing scattering events, is no longer uniform, as is the case for single scattering. Instead, the correlation function measured in an experiment is determined by contributions from all path lengths $s$ weighted by the path length distribution $P(s)$, as 

\begin{equation}
\label{g1_t_as_function_of_P_s}
g_{1}(t)=\int_0^{\infty} {P(s) e^{-\frac{{k_{0}}^2}{3} \MSD s/\lstar}} ds ,
\end{equation}

where $k_0=2 \pi n / \lambda  $ is the magnitude of the photon wave vector in the sample and $s/\lstar$ reflects the number of randomizing scattering events experienced by a photon with path length $s$.~\citep{Pine:1990bj} The basis for this simple form of Eq.\ref{g1_t_as_function_of_P_s} is that each of the approximately $s/\lstar$ randomizing scattering events contributes to a change of this particular photon path by a squared distance of $\MSD$, leading to a partial decorrelation of $g_1(t)$. The cumulative decorrelations from all these randomizing scattering events thus lead to the functional form in Eq.\ref{g1_t_as_function_of_P_s}.
Knowledge on the path length distribution $P(s)$ is therefore essential in the analysis of DWS measurements; without such knowledge the measured correlation functions cannot be related to the dynamics of the scatterers. The path length distribution depends sensitively on the geometry of the sample cell used in the experiment. For sample cells in the shape of a flat slab, infinitely extended in both height and width, $P(s)$ can be expressed in analytical form as a function of $\lstar$ and the thickness $L$ of the sample cell.~\citep{Pine:1990bj,Kaplan:1993cj}\\
This is one of the main reasons why DWS measurements have typically relied on measurements performed in dedicated instruments, employing flat sample cells. 

Such dedicated DWS instruments can also offer other important advantages, in particular for measurements on solid-like, nonergodic samples, where the measured, time-averaged correlation functions are not representative of the ensemble-averaged dynamics of the sample.~\citep{Pusey:1989gy}\\
Methods for acquiring ensemble-averaged correlation functions in DWS measurements include the use of double-cell techniques, where either an ergodic sample with slow dynamics~\citep{Scheffold:2001dj} or a slowly rotating opaque disc~\citep{Zakharov:2006jy} is placed in front of the sample cell. Both these techniques create a slow randomization of the incoming photon paths, resulting in an ensemble-averaging of the collected temporal correlation functions. Either translations of the sample cell or rotations of an opaque disc can also be employed for ensemble-averaging using echo techniques,~\citep{Pham:2004kga} yielding ensemble-averaged correlation functions at long time scales and with excellent statistics.~\citep{Zakharov:2006jy,Reufer:2014fm} While these ensemble-averaging techniques could in principle also be incorporated into a standard goniometer setup, we choose an alternative method, so-called \emph{Pusey-averaging}. This method uses the measured time-averaged correlation function and the measured ensemble-averaged scattered intensity together with a simple theoretical treatment to provide the ensemble-averaged correlation function.~\citep{Pusey:1994wj,Joosten:PhysicalReviewA:1990} 

The ensemble averaged scattering intensity $\left<I\right>_{\mathrm{e}}$ can be readily acquired in separate intensity measurements during which the sample is rotated; the measured dynamics is perturbed by the motion of the sample, but the average scattering intensity is still properly ensemble-averaged. Using the ratio of the ensemble-averaged to the time-averaged scattering intensities $Y= \frac{\left<I\right>_{\mathrm{e}}}{\left<I\right>_{\mathrm{t}}}$, the ensemble-averaged field autocorrelation function $g_1(t)$ can then be estimated as a function of the time-averaged correlation function as
\begin{equation}
\label{Pusey_g1}
g_1(t) = \frac{Y-1}{Y}+\frac{1}{Y} \left[ \tilde{g}_2(t) -\sigma^2 \right]^{\frac{1}{2}} ,
\end{equation}
where $\tilde{g}_2(t)=1+\frac{g_2(t) -1}{\beta}$ is the time-averaged intensity autocorrelation function normalized by the coherence factor $\beta$, which is obtained from the separate ensemble-averaged measurements, and $\sigma^2 = \tilde{g}_2(0)-1$ characterizes the short-time intercept of $\tilde{g}_2(t)$.

We can now use the resulting ensemble-averaged field autocorrelation function $g_1(t)$ to extract viscoelastic properties of the sample, using the microrheology concept ~\citep{Mason:1995jh,nr641}.
To do so, we employ the local power-law approximation developed by Mason et al ~\citep{Mason:2000wz,Dasgupta:PhysRevE:2002}.
In brief, the method is based on the assumption that the Stokes-Einstein relation, which links the thermal motion of particles in a Newtonian liquid to the viscosity of the surrounding liquid, can be generalized to viscoelastic materials with frequency-dependent linear viscoelastic moduli. The approximation also neglects inertial effects on the motion of the probe particles, which is justified for most soft materials at frequencies below $\approx 1\ \mathrm{MHz}$. 

By describing the time-dependent mean square displacement as a local power-law around each data point, the magnitude of the frequency-dependent complex modulus can be expressed in analytical form as

\begin{equation}
\label{Gstar}
|G^\star(\omega)| \approx \frac{k_{\mathrm{B}} T}{\pi a \left<\Delta r^2(1/\omega)\right> \Gamma(1+\alpha(1/\omega))} ,
\end{equation}

where $a$ is the particle radius, $k_{\mathrm{B}}T$ the thermal energy, $\alpha(t)=\frac{\partial \ln(\MSD)}{\partial \ln(t)}$ is the logarithmic slope of the mean-square displacement as a function of lag time, and $\Gamma$ denotes the gamma function. 

\subsection{Sample preparation}
\label{samplepreparation}

Polystyrene particles (micromod Partikeltechnologie GmbH, Germany) coated with a grafted layer ($M_{w}$ = 300 g\slash mol) of poly(ethylene glycol) were used as tracer particles in the DWS measurements. The diameter of the particles is 1$~\mu$m and they are provided suspended in water at a concentration of 5 wt\%. The test samples with tracers in water are prepared by mixing the stock particle suspension with deionized water (Milli-Q water, $\sigma > 18 \mathrm{M}\Omega\cdot \mathrm{cm}$ at 25 °C), to obtain the desired tracer particle concentrations $C_{\mathrm{tracer}}$. To study the effect of the transport mean free path $\lstar$, which is expected to scale as $\lstar \propto 1/C_{\mathrm{tracer}}$, we prepare a series of samples with particle concentrations ranging from $C_{\mathrm{tracer}} \approx 0.3~\mathrm{wt}\%$ to $5~\mathrm{wt}\%$.

The aqueous gelatin gel is prepared by mixing water with 5 wt\% gelatin powder (type A, from porcine skin, Sigma, USA) and 1.25 wt\% of tracer particles at elevated temperatures of $\approx 60^{\circ}\mathrm{C}$. The mixture is homogenized for 30 minutes using a magnetic stirrer, transfered to the cylindrical sample cell used in the experiment, and subsequently allowed to cool down to room temperature. 

\subsection{Light scattering experiments}
\label{lightscatteringexperiments}

All dynamic light scattering experiments are performed in a static and dynamic light scattering setup (ALV CGS--3, ALV GmbH, Germany), equipped with a 50 mW solid state laser ($\lambda = 532\ \mathrm{nm}$) and a goniometer that allows for variation of the detection angle from $\theta\approx 20\ \mathrm{deg}$ to $\theta\approx 160\ \mathrm{deg}$. Measurements are performed in cylindrical cells with outer diameter 10$~\mathrm{mm}$ and inner diameter 8.65$~\mathrm{mm}$; the cell radius relevant to the propagation of photons in the sample cell (see Fig.\ref{Fig1}(C)) is thus $R \approx 4.33~\mathrm{mm}$. Measurements of 30$~\mathrm{s}$ duration are performed at detection angles between $30\ \mathrm{deg}$ and $150\ \mathrm{deg}$ in steps of $10\ \mathrm{deg}$. To minimize the detection of stray light, reflected from surfaces in the setup, our measurements are performed in vertical-horizontal mode, with the incoming light vertically polarized and a horizontal polarizing filter placed in front of the detector. 

For the gelatin samples we use separate experiments on the same sample to determine the ensemble-averaged scattering intensities $\left<I\right>_{\mathrm{e}}$ as well as the coherence factor $\beta$ needed for the Pusey averaging method. In these separate experiments the sample cells are slowly rotated during data acquisition; we perform three such measurements at each scattering angle, each lasting 10 seconds.

In principle, the light scattering measurements we describe in this paper could be performed with any standard goniometer setup; however, not all laser sources that are included in standard goniometer setups may be suitable for performing DWS measurements. In particular, as a result of the long path lengths of the photons through the sample, DWS requires a laser source with a sufficiently long coherence length. Information on the coherence length of a laser is difficult to obtain from the standard information provided by manufacturers, and its measurement requires complex setups only available in specialized optics laboratories.
Within the range of concentrations studied here, the longest relevant path lengths of photons through the samples were limited to around 3 meters. This would be a problem for instance if a Helium-Neon laser were used, which has a typical coherence length of only around 20 cm. Semiconductor lasers, and lasers coupled into single mode fibers, however, typically have much longer coherence lengths that can reach hundreds of meters.
While we have not directly measured the coherence length of our laser source, the good agreement of our path length simulations with experiments(see Results section) makes us confident that in our setup the coherence length of the laser is larger than the relevant path lengths of photons traveling through the sample.

\section{Results and Discussion}
\label{resultsanddiscussion}

\begin{figure*}
\includegraphics[width=\textwidth]{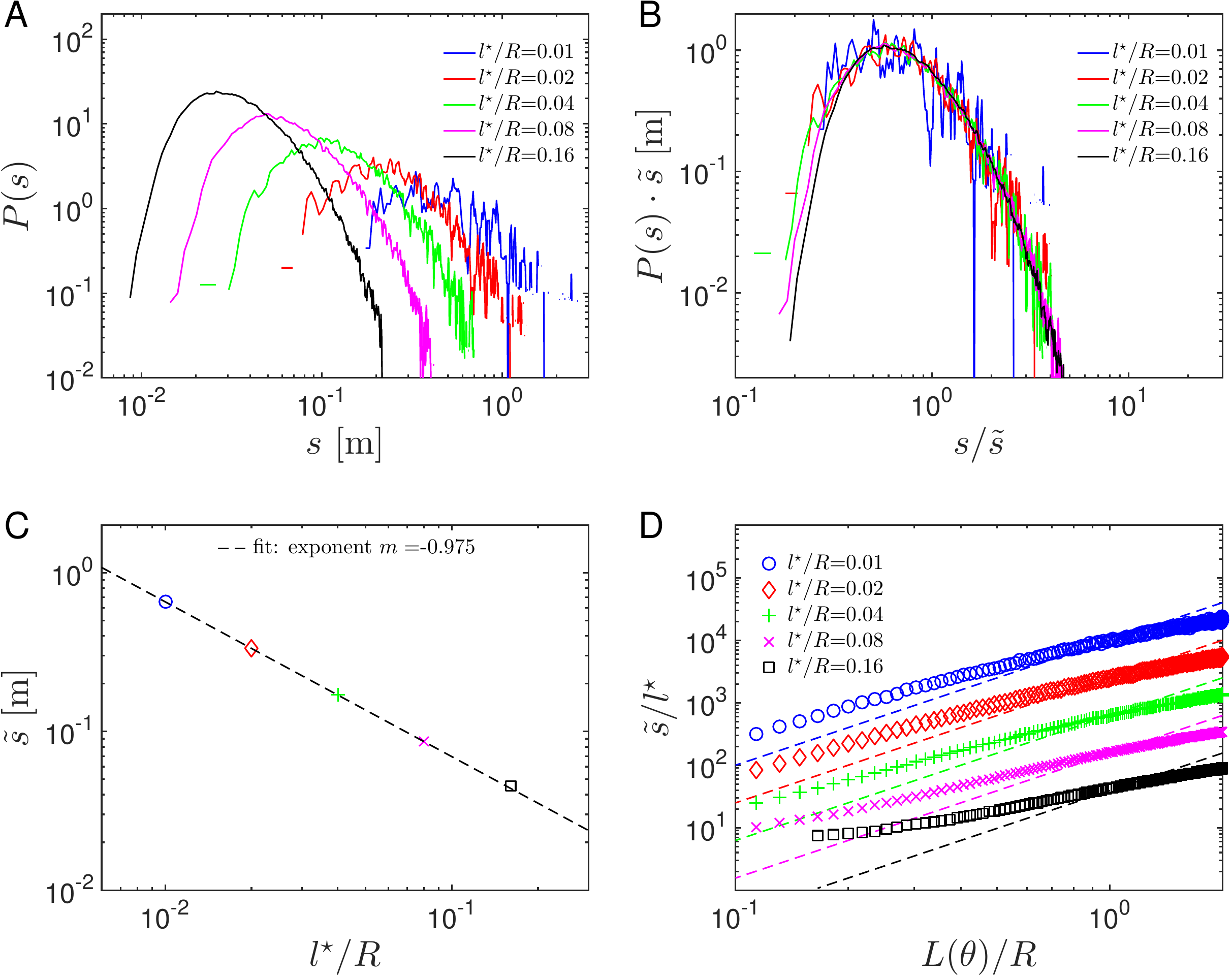}
\caption{(Color online) Simulation results. \textbf{(A)} Path length distribution $P(s)$ for different values of $\lstar/R$, calculated at a fixed detection angle $\theta=90\ \mathrm{deg}$. \textbf{(B)} Master curve of scaled path length distributions, showing that the shapes of $P(s)$ calculated at different $\lstar/R$ are similar. \textbf{(C)} Corresponding average path lengths $\tilde{s}$ as a function of $\lstar/R$. As expected, we find a scaling $\tilde{s} \propto 1/\lstar$, as indicated by a power-law fit to the data (dashed line), yielding an exponent $m = -0.975 \pm 0.05$. \textbf{(D)} Average number of scattering events $\tilde{s}/\lstar$ as a function of the distance $L(\theta)$ between the entry and exit points of the detected photons. The dashed lines serve as a visual reference, indicating the scaling $\tilde{s}/\lstar \propto L(\theta)^2$ that would be expected for an unrestricted random walk.}
\label{Fig2}
\end{figure*}

\subsection{Simulation of photon paths through the sample}
\label{simulationofphotonpathsthroughthesample}

To properly interpret experimental data in a setup with a cylindrical cell, the path length distribution $P(s)$ of photons traveling through the sample is required both as a function of the detection angle $\theta$ and the transport mean free path $\lstar$.

To achieve this, we perform numerical simulations of photons traveling through a cylindrical cell, assuming that they undergo an ideal random walk with step size $\lstar$. In the 2-dimensional coordinate system given in Fig.\ref{Fig1}(C), photons are released at point ($x/R=-1+\lstar/R, y/R=0$), where $\lstar$ is the transport mean free path and $R$ is the radius of the cylindrical cell. Subsequently, each photon is propagated in steps of $\lstar/R$, where each step proceeds in a random (3D)-direction. At the point where the photon exits the cell ( $x^2+y^2>R^2$ ), we evaluate the number $M$ of scattering events, and record the detection event with respect to the observed detection angle $\theta$. 

We do this by dividing the surface of the cylindrical cell into $n_{\mathrm{bins}}$ angular bins, spanning from 0 to 180$\ \mathrm{deg}$ (taking into account the symmetry around the $x$-axis).
In addition, to take into account the 3-dimensional nature of photon transport in the real geometry, we consider the fact that each realization of a 2-dimensional photon path represents a whole range of possible 3-dimensional paths with an identical number of scattering events $M$ and identical ($x$, $y$)-paths. Since in the $z$-direction the photon also performs a (1-dimensional) random walk, we can readily express the probability distribution $p(z)$ for the photon to end up at a position $z$ after propagating $M$ random steps. What is relevant here is the fraction of those paths that will reach the detector, as illustrated in Fig.\ref{Fig1}(D). We assume that all photons with $|z| < \Delta z$ are detected; in accord with the resolution of the angular bins, we set $\Delta z = \frac{\pi}{2 n_{\mathrm{bins}}}$. Then, the fraction $\alpha_{z}$ of contributing 3-dimensional paths is given as

\begin{equation}
\label{fraction_of_contributing_paths}
\alpha_{z} = \mathrm{erf}(\frac{3 \Delta z}{\lstar \sqrt{M}}) ,
\end{equation}

where $\mathrm{erf}$ is the error function, and $\sqrt{\lstar^{2}/3}$ is the effective 1-dimensional step size in the $z$-direction.
We thus account for diffusion in the $z$-direction in our statistics of path length distributions by, instead of adding 1, adding a contribution $\alpha_{z}$ to the angular bin corresponding to each simulated photon path: $f(n_{\mathrm{bin}}) \rightarrow f(n_{\mathrm{bin}}) + \alpha_{\mathrm{z}}$.

Each bin thus represents a detection area of surface area $A_{\mathrm{bin}} \approx \left(\frac{\pi R}{n_{\mathrm{bin}}}\right)^2$. The cumulative value of each angular bin, after propagating $N$ photons and normalizing by $N$, thus defines a (dimensionless) scattering intensity as $I_\mathrm{sim} := \frac{f(n_{\mathrm{bin}} ) }{N}$, representing the probability for a photon to reach the detection area corresponding to bin number $n_{\mathrm{bin}}$. 

In addition to recording the angle where the photons end up, we also record, for each angle $\theta(n_{\mathrm{bin}})$, a distribution of the number of scattering events, by adding a contribution $\alpha_{z}$ to a bin accounting for the number of scattering events at each angle $\theta(n_{\mathrm{bin}})$. The bins are linearly spaced, with bin number 100 representing a number of approximately $\left(L(\theta)/{\lstar}\right)^2$ scattering events, which corresponds to an expected average number of scattering events for a distance $L(\theta)$ between the entry point and the detection point of the photons.\footnote{In the cylindrical cells studied here, the average pathlengths $\tilde{s}$ depend on the detection angle and are typically shorter than estimated from $\tilde{s}/\lstar \approx \left(L(\theta)/{\lstar}\right)^2$, as seen in Fig.\ref{Fig2}(B). } We use 300 bins per angle $\theta(n_{\mathrm{bin}})$, thus accounting for up to 3 times the expected typical number of scattering events; higher numbers, while not counted, in practice are extremely rare in our simulations and do not significantly affect the resulting path length distributions. 

In order to achieve good statistics in the calculated path length distributions, the paths for a large number of photons have to be simulated.

In the actual experiments obtaining good statistics is usually not a problem, due to the enormous number of photons that are propagated.
In our experiments we use a laser with 50$~\mathrm{mW}$ of power at a wave length of 532$~\mathrm{nm}$; this corresponds to $\approx 10^{17}$ photons entering the sample cell, every second. Such numbers are beyond the capability of computer simulations; in comparison, for our calculations we typically simulate $10^{9}$ photons propagating through the sample, which is enough to yield reasonable statistics, and relatively smooth calculated path length distributions.

\begin{figure*}
\includegraphics[width=\textwidth]{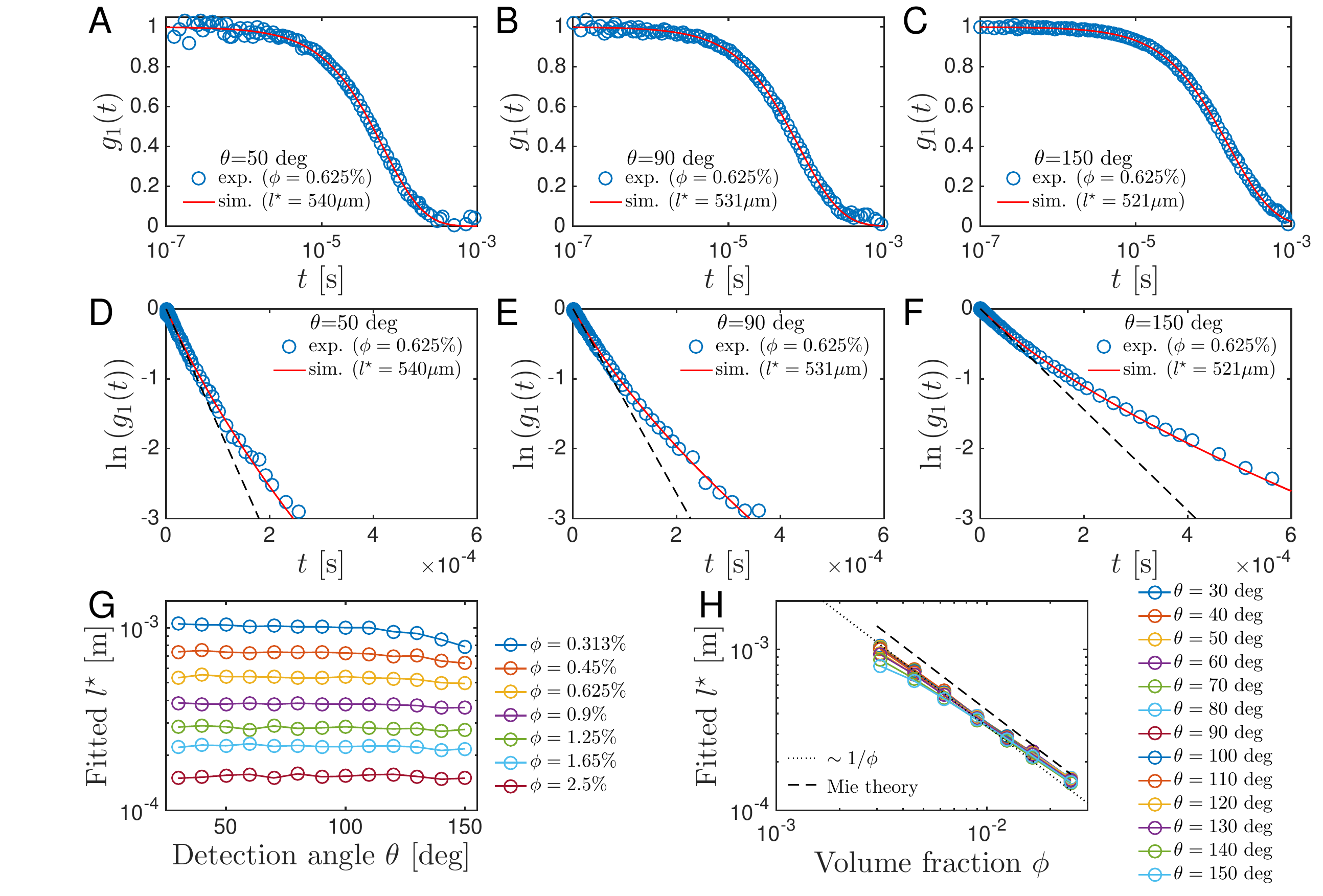}
\caption{(Color online) Simulation results versus experiments for colloidal particles in water ($\phi=0.625\%$, $d=1\mu$m):
Measured (blue circles) and calculated (red line) field correlation function $g_{1}(t)$ for a detection angle of $\theta = 50\ \mathrm{deg}$ \textbf{(A)}, $\theta = 90\ \mathrm{deg}$ \textbf{(B)}, and $\theta = 130\ \mathrm{deg}$ \textbf{(C)}. $\ln(g_{1}(t))$ as a function of $t$ for the same values of $\theta = 50\ \mathrm{deg}$ \textbf{(D)}, $\theta = 90\ \mathrm{deg}$ \textbf{(E)}, and $\theta = 130\ \mathrm{deg}$ \textbf{(F)}. \textbf{(G)}: $\lstar$ obtained from fitting simulation data to experiments, as a function of $\theta$ for different $\phi$. \textbf{(H)}: The same $\lstar$ values plotted as a function of $\phi$ for different values of the detection angle $\theta$; we find approximately $\lstar \propto 1/\phi$, as illustrated by the dotted line. The data is also in fair agreement with predictions from Mie scattering theory,\cite{LSI_website,ochoa2004phdthesis,RojasOchoa:2002ct} shown as a dashed line.}
\label{Fig3}
\end{figure*}

\subsection{Scaling properties of $P(s)$}
\label{scalingpropertiesofps}

Typical obtained simulation results for $P(s)$ are shown in Fig.\ref{Fig2}(A); these curves are calculated at a fixed angle $\theta=90\ \mathrm{deg}$ for different values of $\lstar/R$.
Interestingly, while the average path length decreases with increasing $\lstar$, the shapes of these path length distributions appear surprisingly similar, .

In fact, we can overlay the curves from Fig.~\ref{Fig2}(A) and create a master curve, as shown in Fig.~\ref{Fig2}(B). To obtain this master curve, we have rescaled the path length with a factor $\tilde{s}$ and multiplied the magnitude with the same factor; it turns out that $\tilde{s}$ is the average path length, defined below in Eq.~\ref{s_average}.

Any practical use of the calculated path length distributions requires that $P(s)$ data are available for any arbitrary value of $\lstar$. To address this problem, we calculate path length distributions for different values of $\lstar/R$ and examine the scaling properties of these path length distributions. 

In contrast to an ideal random walk, the path of photons through our cylindrical sample is constricted by the geometry. Nevertheless, the essential scaling properties of a random walk still hold approximately for the path length distributions simulated here. In particular, for an \emph{unrestricted} random walk of step size $\lstar$, we expect the mean square displacement $\left<\Delta R^2\right>$ to be given as $\left<\Delta R^2\right> = M \lstar^2$, with $M$ the number of steps. Considering the average path length 

\begin{equation}
\label{s_average}
\tilde{s}:=\int_0^\infty{s  P(s)}ds\ ,
\end{equation}

we can estimate the average number of scattering events $\tilde{M}$ to travel to a point at distance $L(\theta)$ from the origin to be approximately given as $\tilde{M} \approx \left(L(\theta)/\lstar\right)^2 $, as would be the case for a completely unrestricted random walk.

As the path length is $s=M \lstar$, the average path length should scale with $L(\theta)$ and $\lstar$ as $\tilde{s} \approx L(\theta)^2/\lstar $. Conversely, at fixed detection angle $\theta$ and thus fixed distance $L(\theta)$, we would clearly expect a scaling of $\tilde{s} \propto 1 / \lstar$.

To test this scaling, we examine the $P(s)$ data with respect to both $\lstar$ and the detection angle $\theta$, where a variation of the latter corresponds to a variation of the distance $L(\theta)$ between the entry and detection points of the photons.
In Fig.\ref{Fig2}(C) we plot the average path length $\tilde{s}$ as a function of $\lstar/R$, for simulation data calculated at a single detection angle $\theta=90\ \mathrm{deg}$. Indeed, the data is in excellent agreement with a scaling of $\tilde{s} \propto \lstar^{-1}$; the dashed line in Fig.\ref{Fig2}(C) shows a power-law fit to the data, yielding an exponent of $-0.975 \pm 0.05$. This scaling is a consequence of the self-similarity of random walks, which enables us to approximate each random walk with a ``coarse grained'' version of larger step size; this scaled random walk essentially follows the same path, but, as a result of the increased step size, exhibits a reduced contour length.

In contrast to this simple scaling as a function of $\lstar$, if we examine the average number of scattering events as a function of $L(\theta)$, we find significant deviations from the na\"{i}vely expected scaling $\tilde{s}/\lstar \propto L(\theta)^2$, as shown in Fig.\ref{Fig2}(D). The symbols in this figure show the simulation data for different values of fixed $\lstar/R$, and the solid lines show the simple prediction discussed above, a power-law with exponent 2.
In hindsight, it is clear that such deviations should be expected, as, in contrast to the $\lstar$-dependence at fixed detection angle, a variation of $L(\theta)$ implies a significant modification of the effective sample geometry. It is thus evident that calculations of $P(s)$ at different detection angles are necessary. However, the simple scaling properties with respect to $\lstar$, highlighted in Fig.\ref{Fig2}(B), can be exploited to obtain accurate path length distributions for arbitrary $\lstar$-values, based on simulations performed at a single value of $\lstar/R$.

\begin{figure*}
\includegraphics[width=\linewidth]{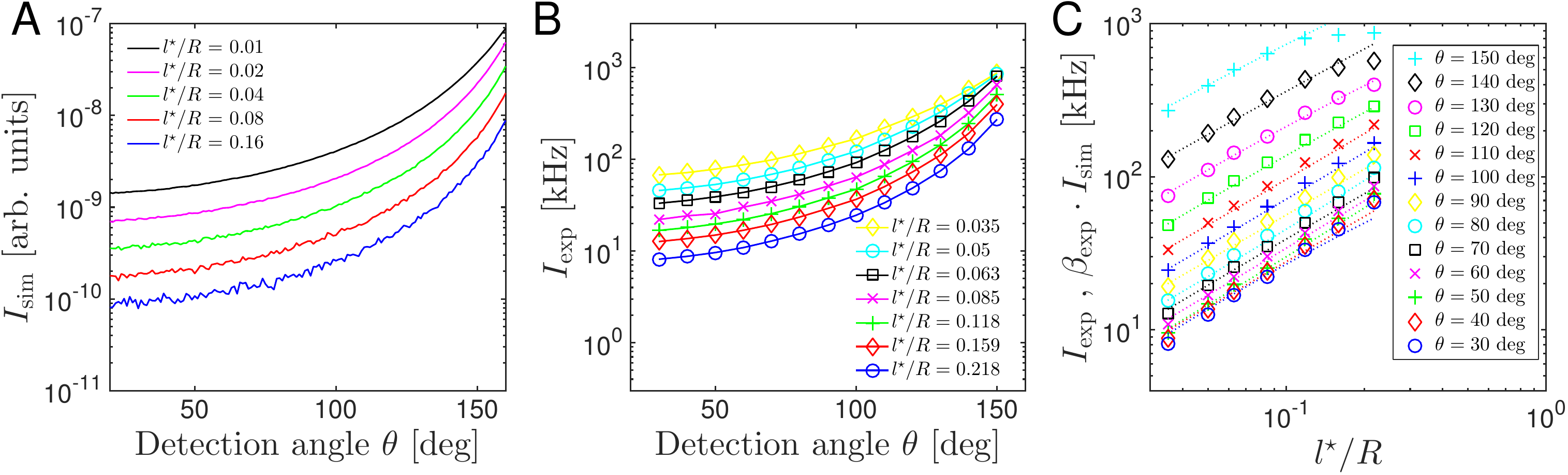}  
\caption{(Color online) Obtaining $\lstar$ from intensity measurements: Calculated and measured intensities as a function of $\theta$ and $\lstar$. \textbf{(A)} Intensity as a function of $\theta$ for simulations employing different values of $\lstar$. \textbf{(B)} Concentration dependence of the measured scattering intensity (for different angles $\theta$). \textbf{(C)} Comparison between experimental and simulation results; scaled by a single constant $\beta_{\mathrm{exp}}$ good agreement between simulation and theory is obtained at all concentrations studied. Thus, once $\beta_{\mathrm{exp}}$ is determined, the measured transport mean free path $\lstar$ can be directly deduced from $I(\theta)$.}
\label{Fig4}
\end{figure*}

\subsection{Determination of $\lstar$ for samples with known tracer dynamics}
\label{determinationoflstarforsampleswithknowntracerdynamics}

Typically, when DWS experiments are used to perform microrheology measurements, uniformly sized tracer particles are added to the soft material of interest. If the transport mean free path $\lstar$ is known, the dynamics of these particles can be extracted, yielding direct information on the viscoelastic properties of the surrounding material. Vice versa, if the viscoelastic properties of the surrounding material are \emph{a priori} known, then $\lstar$ can be extracted from a DWS measurement. A requirement is that the particles scatter much more strongly than the material, ensuring that any detected dynamics are related only to the particle dynamics, and not to fluctuations within the surrounding sample. If this criterion is fulfilled, the simplest method for determining $\lstar$ is to measure the scattering of uniformly sized, spherical particles, suspended in a Newtonian background liquid of known viscosity. 

To test our approach, and for calibration of the transport mean free path $\lstar$, we here perform a series of experiments using samples with different concentrations of uniformly sized polystyrene particles (coated with poly-ethylene glycol, $M_{\mathrm{w}}\approx 300\ \mathrm{g/mol}$, 1 $\mu$m diameter, purchased from micromod GmbH, Germany) suspended in water.
We measure the field autocorrelation function $g_1(t)$ of the scattered light for these samples and test how the correlation functions predicted from our photon path simulations compare to these experimental data. As shown in Fig.\ref{Fig3}(A-C), where we show data on a suspension of particles at a volume fraction $\phi=0.625\%$ and detection angles $\theta=50\ \mathrm{deg}$, $90\ \mathrm{deg}$, and $130\ \mathrm{deg}$, we obtain remarkably good agreement between experiments (shown as blue circles) and simulations (shown as red lines), where $\lstar$ is the only adjustable parameter. While the dynamics is expected to be purely Brownian, with $\MSD$ increasing linearly with time $t$, due to the broad path length distribution of photons passing through the sample, $g_1(t)$ deviates significantly from the single exponential decay that would be observed in single scattering experiments. This can be more clearly seen in Fig.\ref{Fig3}(D-F), where $\ln(g_1(t))$ is plotted as a function of time; in such a plot an exponential decay would appear as a straight line, as illustrated by the dashed lines, which show exponential fits to the short-time regime of $g_1(t)$.
The non-exponential shape of the data thus becomes evident and is captured very well by the curves predicted from our simulations, shown as red lines.
Fits performed for the same sample, but at different angles, should yield the same $\lstar$-values. Indeed, we obtain good agreement between the $\lstar$-values extracted from the data in Fig.\ref{Fig3}(A-C): we obtain $\lstar=540\ \mu\mathrm{m}$, $\lstar=533\ \mu\mathrm{m}$, and $\lstar=523\ \mu\mathrm{m}$ at angles of $\theta=50\ \mathrm{deg}$, $90\ \mathrm{deg}$, and $130\ \mathrm{deg}$, respectively.
In fact, we obtain good agreement between measurements taken at different angles for all the concentrations studied, with volume fractions ranging from $\phi=0.313\%$ to $2.5\%$. As shown in Fig.\ref{Fig3}(G), the fitted $\lstar$-values as a function of $\theta$ exhibit only small variations. Somewhat larger deviations are observed for the sample with the lowest concentration, at the largest detection angles. We attribute this to the fact that this sample has the longest $\lstar$, combined with the shortest distances $L(\theta)$ between entry point and exit point of the photons; $\lstar \approx 1\mathrm{\ mm}$ and $L(\theta) \approx 2.2\ \mathrm{mm}$ and thus $L(\theta)/\lstar \approx 2.2 $. In this case the path length of photons is no longer adequately described as an ideal random walk.\\
Besides these discrepancies at small values of $L(\theta)$, the fitted $\lstar$-values depend only on the volume fraction, irrespective of the detection angle. To examine the $\phi$-dependence of the data, in Fig.\ref{Fig3}(H) we plot $\lstar$ as a function of $\phi$, observing approximately the expected scaling $\lstar \propto 1/\phi$,~\citep{Durian:1995hp} as indicated by the dotted line. The data is also in fair agreement with Mie scattering calculations plotted as a dashed line in Fig.\ref{Fig3}(H).~\citep{ochoa2004phdthesis,RojasOchoa:2002ct} The calculations are performed using the web application available on the website of LS Instruments, Switzerland,~\citep{LSI_website} using as input parameters the particle size, the wavelength of the laser $\lambda=532\ \mathrm{nm}$, as well as a refractive index of $n_{\mathrm{PS}}=1.598$ for the particles and $n_{\mathrm{H}_2\mathrm{O}}=1.33$ for water.

\subsection{Obtaining $\lstar$ from intensity measurements}
\label{obtaininglstarfromintensitymeasurements}

In the absence of absorption, the intensity detected at each angle should be fully determined by the transport mean free path of photons in the sample.
While absorption is relatively straightforward to include in the current data analysis, here we choose to neglect its effects since absorption is relatively weak in the aqueous samples studied; the typical absorption length is much longer than the typical path length of photons through the samples.
As a result, after calibration using a reference sample of known dynamics, a simple measurement of the scattering intensity at different angles on the sample of interest is sufficient for determining its $\lstar$.
To validate this, we compare the scattering intensities predicted from the simulations with those measured in experiments performed on our polystyrene suspensions.

In Fig.\ref{Fig4}(A) we plot the scattering intensity $I_\mathrm{sim}$ as a function of detection angle $\theta$, as predicted from the photon path simulations. The shape of these curves is very different from those typically obtained in single scattering experiments on dilute suspensions, where generally the intensity is highest at small detection angles, corresponding to low $q-\mathrm{values}$. In the highly multiple scattering regime, however, the intensity is generally highest for detection points closest to the entry point of photons into the sample, which corresponds to large $\theta$-values.

Importantly, we find that the angular dependence of the recorded scattered intensity for the tracer particle suspensions agrees remarkably well with the behavior predicted from our simulations, as shown in Fig.\ref{Fig4}(B). For both simulations and experiments, we observe a ratio of $\approx 40$ between the intensities measured at $\theta=30\ \mathrm{deg}$ and $\theta=150\ \mathrm{deg}$. Moroever, comparing Fig.\ref{Fig4}(B) with Fig.\ref{Fig4}(A), we observe that the shapes of the simulated intensity curves are very similar to those of the experimental data.
In fact, the two data sets can be superposed simply by scaling the simulated curves with one single factor $\beta_\mathrm{exp}$, the value of which depends on experimental parameters such as the size of the detection area, and the distance between the detector and sample. We find that good agreement between the measured and simulated curves is obtained for a value of $\beta_\mathrm{exp} \approx 8 \cdot 10^{10} \mathrm{\ Hz}$, as shown in Fig.\ref{Fig4}(C).

Thus, given $\beta_\mathrm{exp}$, a good estimate of $\lstar$ can be determined for a sample of unknown properties solely by measuring the scattered intensity at different angles.

\begin{figure*} 
\includegraphics[width=\linewidth]{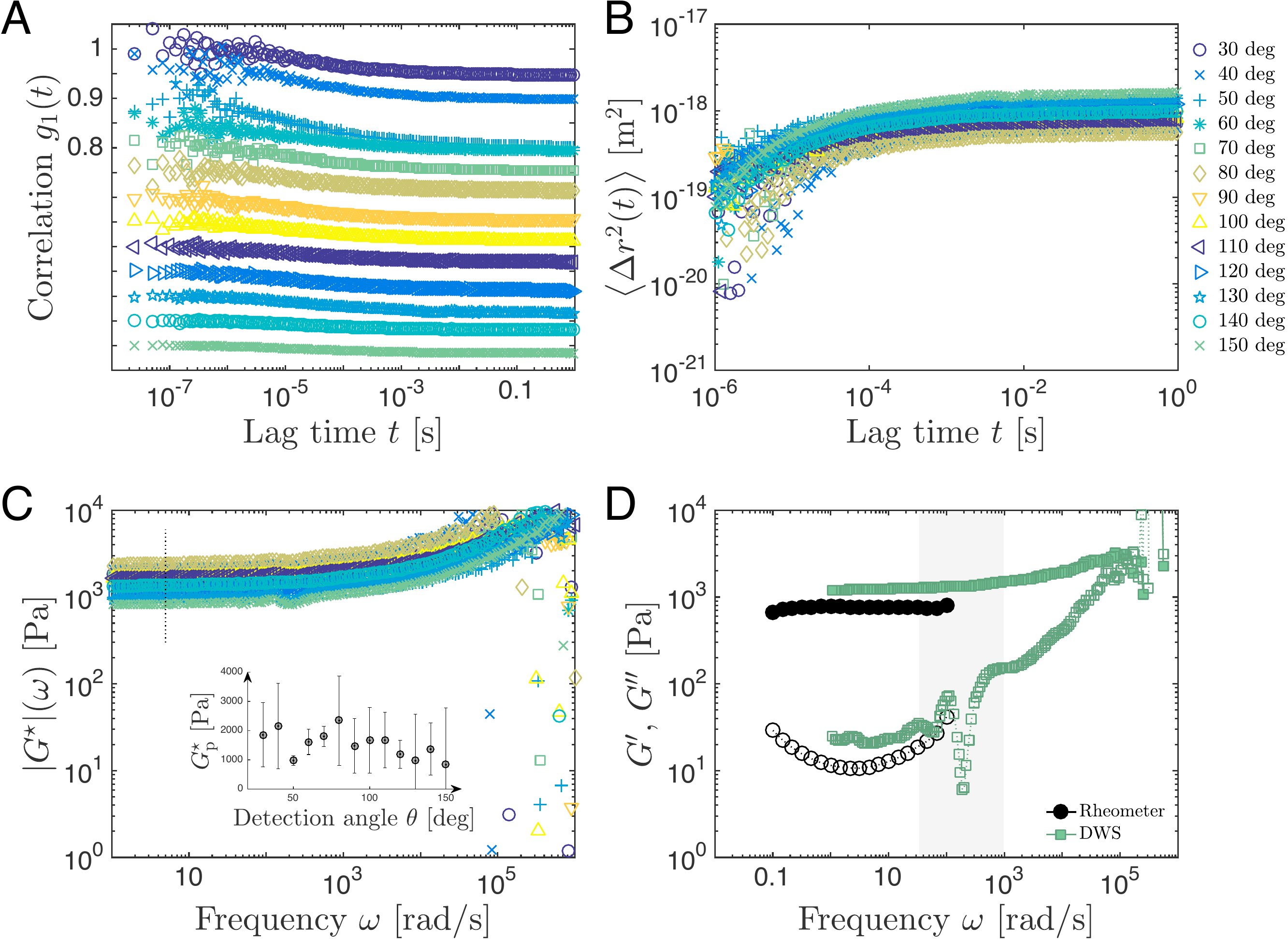}
\caption{Measurements on a solid-like, non-ergodic sample, an aqueous gelatin gel at a concentration of 5 wt\% with embedded tracer particles (1.25 wt\%, $a=1 \ \mu\mathrm{m}$ ). \textbf{(A)}: Pusey-averaged field autocorrelation functions $g_1(t)$ as a function of lag time $t$, measured for detection angles ranging from $\theta=30\ \mathrm{deg}$ to $\theta=150\ \mathrm{deg}$. Curves are offset vertically by increments of -0.05 for clarity. \textbf{(B)}: Mean-square displacements extracted from the same data, plotted as a function of time, yielding fair agreement between measurements taken at different angles. Note that sub-nanometer displacements are accessed. \textbf{(C)} Magnitudes of the corresponding complex shear moduli $|G^{\star}(\omega)|$ as a function of frequency $\omega$. The inset shows plateau values $G^{\star}_p$, accessed at a frequency of 5 rad/s (indicated as a dotted line in the main plot), as a function of detection angle. Within the estimated error bars, no strong trend in the data is observed; averaging over data from different $\theta$ thus appears justified. \textbf{(D)} Storage modulus $G^{\prime}(\omega)$ (solid squares) and loss modulus $G^{\prime\prime}(\omega)$ (open squares) averaged over all measurements taken at different angles as a function of frequency $\omega$. Comparing these averaged moduli to results from conventional oscillatory rheology, with $G^{\prime}(\omega)$ shown as solid black circles and $G^{\prime\prime}(\omega)$ shown as open black circles, we observe very good agreement.}
\label{Fig5}
\end{figure*}

\subsection{Test on a viscoelastic material}
\label{testonaviscoelasticmaterial}

Finally, we test the use of our method in the context of microrheology, where the measured correlation functions and the corresponding tracer particle mean-square displacements $\MSD$ are used for the determining viscoelastic properties of a sample.
As a test material we use a common solid-like soft material, an aqueous gelatin gel at a concentration of 5 wt\%. The plateau storage modulus of this material is on the order of 1 kPa, which means that for the case of micron-sized tracer particles we need to be able to access particle displacements at sub-nanometer length scales. DWS is ideally suited for this, since the displacements of all the tracer particles encountered by a photon on its path through the sample cumulatively contribute to changing the total photon path length.

We perform measurements on the gelatin sample for detection angles ranging from $\theta=30\ \mathrm{deg}$ to $\theta=150\ \mathrm{deg}$. The sample is highly non-ergodic, as indicated by the fact that the intercept of the measured intensity correlation functions $g_1(t)$ varies significantly between measurements.
We therefore use the Pusey-averaging procedure to obtain a good estimate of the ensemble averaged correlation functions from the measured, time-averaged correlation functions, as outlined in the experimental section. As expected, and shown in Fig.\ref{Fig5}(A), the resulting ensemble-averaged field correlation functions $g_1(t)$ vary with the detection angle $\theta$. These correlation functions do not decay significantly; they reach a plateau at values of $g_1(t) > 0.9$ at the longest time scales accessed in the experiments. This reflects the fact that the gelatin sample has a relatively high modulus and the thermal motion of the tracer particles is therefore limited to short length scales.
Using the procedure outlined above, we obtain the transport mean free path $\lstar$ of the gelatin sample directly from the measured scattered intensities, using the intensity scaling factor $\beta_{\mathrm{exp}}$ as determined from the measurements on our pure tracer suspensions. The mean-square displacements of tracer particles in the gelatin sample are obtained by numerically inverting Eq.\ref{g1_t_as_function_of_P_s}, using the measured $g_1(t)$, $\lstar$, and the calculated $\theta$-dependent path length distribution $P(s)$ as input. The corresponding mean-square displacements are shown in Fig.\ref{Fig5}(B). Given the highly nonergodic nature of the sample studied, the data taken at different detection angles are in fair agreement; note that the magnitudes of the accessed particle displacements are in the sub-nanometer range.
We can now convert these data to viscoelastic moduli, using the microrheology concept ~\citep{Mason:1995jh,nr641} and the local power-law approximation ~\citep{Mason:2000wz,Dasgupta:PhysRevE:2002}, developed by Mason et al. The magnitude of the resulting complex modulus $|G^{\star}(\omega)|$ is on the order of 1 kPa and depends only weakly on frequency, as shown in Fig.\ref{Fig5}(C). The curves obtained for different detection angles exhibit significant variations, as shown in the inset, where we plot the low frequency plateau values $G^{\star}_p = |G^{\star}(\omega=5\ \mathrm{rad/s})|$ as a function of detection angle (the frequency of 5 rad\slash s is indicated as a dotted line in the main plot). Since $|G^{\star}(\omega)|$ is approximately inversely proportional to the mean square displacement $\left<\Delta r^2(t=1/\omega)\right>$, these variations in the magnitude of the complex modulus directly reflect those observed in the tracer mean-square displacements. 

A simple error analysis (see SI) suggests that the main sources of errors are on the one hand \emph{statistical errors} in the intensity correlation function as a result of the finite measurement duration, and on the other hand errors introduced via the Pusey averaging procedure via the \emph{error in the intensity ratio} $Y=\frac{I_t}{I_e}$ between the time-averaged and the ensemble-averaged scattering intensities.\\
The total corresponding relative error in the modulus, $\Delta G / G$, can be expressed as a function of the relative errors in the intensity ratio, $\Delta Y/Y$, and the ratio between the probed time scale $t$ and the measurement duration $T$, as~\citep{degiorgio1971intensity}

\begin{equation}
\label{totalerror}
\frac{\Delta G}{G} \approx \frac{\Delta \MSD}{\MSD} \approx \frac{\Delta Y}{Y} + \frac{3}{{g_1}^2 \ln(g_1)}\sqrt{\frac{t}{T}}
\end{equation}

We can estimate the error in determining the intensity ratio $Y$ from the standard deviation of the 3 ensemble-averaged intensity measurements taken at each angle as $\Delta Y \approx Y \cdot \Delta I_\mathrm{e}/I_\mathrm{e}$, where $\Delta I_\mathrm{e}$ is taken as the standard deviation of the 3 intensity measurements, and $I_{\mathrm{e}}$ is the average ensemble-averaged intensity. The relative error in the mean square displacement $\epsilon_\mathrm{MSD}=\Delta(\MSD)/\MSD$ that results from the error in $Y$ can be estimated as $\epsilon_\mathrm{MSD} \approx {\partial \MSD}/{\partial g_1(t)} \cdot {\partial g_1(t)}/{\partial Y} \cdot {\Delta Y}/{Y}$. As $ \partial g_1(t)/{\partial Y} \approx 1/{Y^2}$ and ${\partial \MSD}/{\partial g_1(t)} \approx {-\MSD}/[ln(g_1(t))\cdot g_1(t)]$, we find $\epsilon_\mathrm{MSD} \approx -\left[\ln(g_1(t)) g_1(t) Y^2\right]^{-1} \frac{\Delta Y}{Y}$. Because the modulus $|G^\star (\omega)|$ is essentially given as the inverse of the mean-square displacement, it has the same typical relative error $\epsilon_{G^\star} \approx \epsilon_\mathrm{MSD}$.
For our measurements we find typical values $\epsilon_{G^\star} \approx 0.5$, as shown in the inset of Fig.\ref{Fig5}(C), where the corresponding errors bars are shown for each angle. 

While the magnitude of these errors is indeed significant, the observed angle-dependent variations remain within a factor of $\approx 2$ and exhibit no clear systematic dependence on the detection angle. The latter indicates that the observed variations are the result of random, rather than systematic measurement errors, and thus do not reflect a systematic issue with the data analysis method.
This suggests that performing an average of the data measured at different detection angles provides a more reliable result than data from a single detection angle measurement.
We thus average the mean-square displacements obtained at all accessed angles, using the inverse of the estimated error bars as weights in the averaging procedure. Using this averaged data to calculate the viscoelastic response of the sample we obtain an averaged viscoelastic response, shown in Fig.~\ref{Fig5}(D), where we plot the storage modulus $G^{\prime}$ (solid squares) and the loss modulus $G^{\prime\prime}$ (open squares) as a function of frequency $\omega$. The oscillations observed in the data at a frequency of $\approx 50-200\mathrm{\ Hz}$ are likely the result of a mechanical disturbance or vibration that we could not eliminate in our experimental light scattering setup on this highly out-of-equilibrium sample. The effect can be directly observed in the measured correlation functions at the corresponding time scales, as seen in Fig.\ref{Fig5}(A). For mechanically weaker samples we have not observed these types of oscillations in our setup. As a result of the importance of the time-derivative of $\MSD$ in determining the viscoelastic moduli, the oscillations in the $g_1(t)$-data are amplified in the corresponding viscoelastic moduli. The frequency range where we do not trust the data as a result of these oscillations is indicated with a grey background in Fig.\ref{Fig5}(D). Nevertheless, besides these oscillations, we obtain very good agreement with measurements performed using a conventional oscillatory rheometer, shown in the same figure as solid black circles for $G^{\prime}$ and open black circles for $G^{\prime\prime}$.

\section{Conclusions}
\label{conclusions}

We have developed a simple method for properly interpreting dynamic light scattering data from highly multiple scattering samples using a standard dynamic light scattering setup with a cylindrical sample geometry.
By performing ideal random walk simulations within a cylindrical geometry, we predict the path length distribution $P(s)$ of photons passing through the sample cell. This enables us to extend the use of DWS measurements to standard dynamic light scattering instruments. The method can be applied in the context of microrheology, where the dynamics of embedded tracer particles are used to access the frequency-dependent viscoelastic response of soft materials. 

The main strength of our approach, besides not requiring a dedicated instrument, lies in the fact that by varying the detection angle we can access a wide range of different effective sample geometries with different average path lengths, using one single cylindrical sample cell. This variation of the detection angle is analogous to performing a series of conventional DWS measurements using a series of sample cells with varying thickness or with varying tracer particle concentrations. 

We have further illustrated the usefulness of our method for DWS-based microrheology on a soft solid, gelatin, for which we obtain a very good agreement with macroscopic oscillatory rheology experiments. Moreover, data recorded for different detection angles enable an important consistency check for the microrheology measurements and our results illustrate that the accuracy of such microrheology measurements can be improved by averaging over measurements obtained at different detection angles.

\begin{acknowledgments}
JM and HMW thank the Royal Society for travel support. The work of FA and HMW forms part of the research programme of the Dutch Polymer Institute (DPI), project \#738. We thank the Institute for Complex Molecular Systems (ICMS) at Eindhoven University of Technology for support.
\end{acknowledgments}

\newpage

\newpage

\onecolumngrid
\appendix

\section*{Supplementary Information for article "Diffusing-Wave Spectroscopy in a Standard Dynamic Light Scattering Setup"}

\subsection*{Error analysis for DWS and microrheology data}
\label{erroranalysisfordwsandmicrorheologydata}

In the following we provide a brief analysis aimed at estimating the relative experimental errors associated with our DWS-based microrheology measurements.\\
In DWS, the dynamics of tracer particles is quantified in terms of temporal autocorrelation functions, from which the time-dependent mean-square displacement ($\left<\Delta r^2(t)\right>$, hereafter abbreviated as MSD($t$) ) of the particles can be calculated. Using a generalized Stokes-Einstein relation, the MSD's are then converted to viscoelastic moduli, with the magnitude of the complex modulus $|G^\star(\omega)|$ approximately proportional to the inverse of the mean-square displacement at a time scale $t=1/\omega$.\\
The relative error in the modulus, $\Delta G^\star / G^\star$, is therefore in good approximation the same as the relative error in the mean-square displacement $\Delta \mathrm{MSD} /\mathrm{MSD}$. We identify two main sources of error that ultimately determine this relative error in the measured mechanical response:\\
1. \emph{Statistical errors} in the intensity correlation function as a result of the finite measurement duration.\\
2. For nonergodic samples, additional errors are introduced via the Pusey averaging procedure used to estimate ensemble-averaged temporal autocorrelation functions. These errors are introduced via the \emph{relative error in the intensity ratio} $Y=\frac{I_t}{I_e}$ between the time-averaged and the ensemble-averaged scattering intensities.

\subsubsection{Statistical errors in the intensity correlation function}
\label{statisticalerrorsintheintensitycorrelationfunction}

Given a measurement duration $T$, the statistical error in the intensity correlation function $g_2(t)$ at lag time $t$ can be expressed as~\citep{degiorgio1971intensity}

\begin{equation}
\label{g2error}
\Delta g_2(t) \approx 6 \sqrt{\frac{t}{T}} .
\end{equation}

Given the Siegert relation $g_1(t) = \sqrt{g_2(t) - 1}$, this translates to an error in the field correlation function $g_1(t)$ as 

\begin{equation}
\label{g1error}
\Delta g_1(t) \approx \frac{3}{g_1(t)} \sqrt{\frac{t}{T}}  .
\end{equation}

For the purposes of this error analysis, we assume a simplified relationship between the field correlation function $g_1(t)$ and the mean-square displacement $\mathrm{MSD}(t)$, $g_1(t) \approx e^{-k_0^2/3 \frac{ \tilde{s}}{l^\star} \mathrm{MSD}(t)}$, with $k_0$ the wave vector of the laser light, $\tilde{s}$ the average path length of photons, and $l^\star$ the transport mean free path. 

We can then express the mean square displacement as $\mathrm{MSD}(t) \approx 3 \lstar \ln(g_1(t) / \left({k_0}^2 \tilde{s}\right)$.
With $\partial \mathrm{MSD} / \partial g_1 \approx \frac{3 \lstar}{{k_0}^2 \tilde{s}} \frac{1}{g_1} \approx \mathrm{MSD} / (g_1 \ln(g_1))$ and $\Delta g_1 \approx \frac{3}{g_1}\sqrt{\frac{t}{T}}$ we can now express the total relative error in the complex modulus as

\begin{equation}
\label{stat_errors}
\Delta G^\star / G^\star \approx \Delta \mathrm{MSD} / \mathrm{MSD} \approx \frac{1}{\mathrm{MSD}}\frac{\partial \mathrm{MSD}}{\partial g_1} \Delta g_1 \approx \frac{1}{g_1\ln(g_1)}\frac{3}{g_1} \sqrt{\frac{t}{T}}
\end{equation}

\subsubsection{Errors introduced during the Pusey averaging procedure}
\label{errorsintroducedduringthepuseyaveragingprocedure}

For nonergodic samples, additional errors are introduced via the Pusey averaging procedure used to estimate ensemble-averaged temporal autocorrelation functions. These errors are introduced via the \emph{relative error in the intensity ratio} $Y=\frac{I_t}{I_e}$ between the time-averaged and the ensemble-averaged scattering intensities.
The resulting error in the mean square displacement is
\begin{equation}
\label{pusey1}
\Delta \mathrm{MSD} \approx \frac{\partial \mathrm{MSD}}{\partial g_1} \frac{\partial g_1}{\partial Y} \Delta Y \ ,
\end{equation}
where $\frac{\partial \mathrm{MSD}}{\partial g_1} \approx \frac{\mathrm{MSD}}{g_1 \ln(g_1)}$, as derived above.\\
To arrive at an expression for the second term in Eq.\ref{pusey1}, we write down the relationship between the field correlation function $g_1(t)$ and the intensity ratio $Y$ as explained in the main manuscript,
\begin{equation}
\label{g1vsy}
g_1(t) = \frac{Y-1}{Y}+\frac{1}{Y}\left[ \tilde{g}_2(t) - \sigma^2 \right]^{\frac{1}{2}} \ ,
\end{equation}
where $\tilde{g}_2(t) = 1 + \frac{g_2(t)-1}{\beta}$ is the time-averaged intensity autocorrelation function normalized by the coherence factor $\beta$, and $\sigma^2 = \tilde{g}_2(t) 1$ characterizes the short-time intercept of $\tilde{g}_2(t)$. This yields the second term in Eq.\ref{pusey1} as $ \frac{\partial g_1}{\partial Y} \approx \frac{1}{Y^2}\left[ 1 - \sqrt{\tilde{g}_2 - \sigma^2} \right]$, and with $\sqrt{\tilde{g}_2 - \sigma^2} \approx 1+ Y g_1 - Y$, this results in
\begin{equation}
\label{dg1dy}
 \frac{\partial g_1}{\partial Y} \approx \frac{1 - g_1}{Y}
\end{equation}

The total relative error introduced by the intensity ratio $Y$ is thus 

\begin{equation}
\label{totalerrorY}
\Delta \mathrm{MSD} \approx \frac{1 - g_1}{g_1 \ln(g_1)} \cdot \frac{\Delta Y}{Y} \ .
\end{equation}

Plotting the function $f(x)=\frac{1-x}{x\ln(x)}$, we observe $|f(x)| \approx 1$ for values of $x$ close to 1. For our solid-like samples, $g_1$ is close to 1 at long times, and thus $\left| \frac{1-g1}{g_1 \ln(g_1)}\right| \approx 1$; therefore in this case we can make the simple approximation $\frac{\Delta\mathrm{MSD}}{\mathrm{MSD}} \approx \frac{\Delta Y}{Y}$.

\subsubsection{Total estimated error}
\label{totalestimatederror}

The total estimated error as a result of the above two main sources of error can then be written as
\begin{equation}
\label{totalerror}
\Delta G^\star / G^\star \approx \Delta \mathrm{MSD} / \mathrm{MSD} \approx \frac{3}{{g_1}^2\ln(g_1)} \sqrt{\frac{t}{T}} + \frac{1 - g_1}{g_1 \ln(g_1)} \cdot \frac{\Delta Y}{Y}
\end{equation}

Of these two contributions to the experimental error, the latter usually dominates, provided that the measurement duration is sufficiently long, such that the factor $\sqrt{\frac{t}{T}}$ becomes small enough. Nevertheless, for $g_1(t)$ sufficiently close to 1, the term $\frac{1}{\ln(g_1)}$ would become very large, and eventually lead to the first term becoming dominant.\\
Indeed, we expect the errors to be highest for cases where $g_1(t)$ is very close to zero (as a result of the factor $\frac{3}{{g_1}^2\ln(g_1)}$ or $\frac{1-g1}{g_1 \ln(g_1)}$, respectively), or very close to unity (as a result of the factor $\frac{1}{\ln(g_1)}$).

\subsection{Code for data analysis of DWS in a cylindrical cell}
\label{codefordataanalysisofdwsinacylindricalcell}

\subsubsection{Full codes (written in Matlab) for analyzing DWS data in a cylindrical cell}
\label{fullcodeswritteninmatlabforanalyzingdwsdatainacylindricalcell}

The complete numerical codes used for analyzing dynamic light scattering data using the approach outlined in the manuscript can be obtained on the author's website $<$www.mate.tue.nl\slash \ensuremath{\sim}wyss$>$ or on request by sending an email to Hans Wyss at \href{mailto:H.M.Wyss@tue.nl}{H.M.Wyss@tue.nl}. Please also address any questions regarding the code and\slash or data analysis to the same email address.\\
The specific code for the random walk simulation for calculating the path length distribution $P(s)$ is listed below. 

\subsubsection{Code for calculating the path length distributions}
\label{codeforcalculatingthepathlengthdistributions}

The main code for calculating the path length distribution $P(s)$ is a simple random walk simulation with step length $l^\star$, as described in the main manuscript. Below is the C-code (and the MEX function called by our main Matlab code) that accomplishes this task; the results are kept track of in the output array $y$, which for each bin corresponding to a segment of detection angle and a segment of path length keeps track of the number of photons exiting the cell within the corresponding angle range and within the corresponding path length range. Angular bins evenly divide the angular space between 0 and 180 degrees; we usually choose 180 bins of 1 degree width. Path length bins are also linearly spaced, with 300 bins total and the 100$^{\mathrm{th}}$ bin corresponding to a path length of $\left(L(\theta)/{\lstar}\right)^2$ in units of $l^\star$, where $L(\theta)$ is the distance between the entry point and the exit point of the simulated photon, as a function of the detection angle $\theta$. 

\begin{center}\rule{3in}{0.4pt}\end{center}

\textbf{Listing of ``pathlengthsCyl\_P\_s.c'':}

\lstset{
  language=C,                
  numbers=left,                   
  stepnumber=1,                   
  numbersep=5pt,                  
  backgroundcolor=\color{white},  
  showspaces=false,               
  showstringspaces=false,         
  showtabs=false,                 
  tabsize=2,                      
  captionpos=b,                   
  breaklines=true,                
  breakatwhitespace=true,         
  title=\lstname,                 
}

\begin{lstlisting}
#include "mex.h"
#include <stdio.h>
#include <stdlib.h>
#include <time.h>
#include <math.h>

/*
 *
 * Function that calculates the pathlength distribution for diffusion of photons
 * through a cylinder
 *
 */

void pathlengthsCyl(double y[], double x[], size_t mrows, size_t ncols)
{
    int ii,m,n,step,anglebin,Nbins,N_iter,index1,index2,index3;
    double xx,yy,zz,dxx,dyy,dzz,lstar,LL,pi,deltaz,weight,dist;
    
    lstar=x[0];
    Nbins=x[1];
    N_iter=x[2];
    pi=3.14159265359;
    
    //srand( (unsigned)time( NULL ) );
    srand ( rand() ^ time(NULL) );
    
    for(m=0;m<mrows;m++)
    {
        for(n=0;n<ncols;n++)
        {
           index1=(m % mrows)+mrows*n;
           y[index1]=0;
           
        }
    }
    
    for(n=0;n<ncols;n++)
    {
        dist=sqrt( (-1-cos(pi/Nbins*n))*(-1-cos(pi/Nbins*n))+sin(pi/Nbins*n)*sin(pi/Nbins*n) );
        index1=2+mrows*n;
        y[index1]=ceil(dist*dist/lstar/lstar/100); //write column 2: the width of each bin (set as one hundreds of the expected average number of scattering events.)
    }
    
    
    for(ii=0;ii<N_iter;ii++)
    {
        xx=-1+lstar;
        yy=0;zz=0;
        step=1;
        while ((xx*xx+yy*yy)<1)
        {
            
            // Random unit vector of length lstar:
            dxx=2;dyy=2;dzz=2;
            while (dxx*dxx+dyy*dyy+dzz*dzz>1) // make sure (dxx,dyy,dzz) is a vector of random direction.
            {
                dxx=2*(double)(rand()) / (double)(RAND_MAX)-1;
                dyy=2*(double)(rand()) / (double)(RAND_MAX)-1;
                dzz=2*(double)(rand()) / (double)(RAND_MAX)-1;
            }
            LL=sqrt(dxx*dxx+dyy*dyy+dzz*dzz);
            dxx=dxx/LL*lstar;dyy=dyy/LL*lstar;dzz=dzz/LL*lstar;
            // Propagate by a step lstar:
            xx=xx+dxx;yy=yy+dyy;zz=zz+dzz;
            step++;
        }
        LL=sqrt(xx*xx+yy*yy);
        xx=xx/LL;yy=yy/LL;zz=zz/LL;
        anglebin=ceil(acos(xx)*(Nbins-1)/pi);
        
        index1=mrows*anglebin;
        deltaz=pi/2/Nbins;
        weight=erf(3*deltaz/lstar/sqrt(step)); //account for probability in z-direction to hit area around zz=0
            if(weight==0) mexPrintf("erf: %f\n",weight);
            y[index1]=y[index1]+weight;
            // Update average number of steps and corresponding added weights:
            index2=(1 % mrows)+mrows*anglebin;
            y[index2]=y[index2]+step*weight;
            index3=floor(step/y[anglebin*mrows+2])+3+mrows*anglebin;
            if(floor(step/y[anglebin*mrows+2])+3<mrows)
            {
                y[index3]=y[index3]+weight; // add value to bin
            }

    }
    
    for(anglebin=0;anglebin<Nbins;anglebin++)
        
    {
        index1=mrows*anglebin;
        index2=(1 % mrows)+mrows*anglebin;
        if(y[index1]==0)
        {
          //mexPrintf("y[index1] is zero. index1:%i y[index1]: %f, y[index2]: %f\n",index1,y[index1],y[index2]);
        }
        if(y[index2]>0)
        {
        y[index2]=y[index2]/y[index1];
        }
    }

}





void mexFunction( int nlhs, mxArray *plhs[],
        int nrhs, const mxArray *prhs[] )
{
    double *x,*y;
    double lstar;
    int Nbins;
    size_t mrows,ncols,mrows_out,ncols_out;
    
    /* Check for proper number of arguments. */
    if(nrhs!=1) {
        mexErrMsgIdAndTxt( "MATLAB:timestwo:invalidNumInputs",
                "One input required.");
    } else if(nlhs>1) {
        mexErrMsgIdAndTxt( "MATLAB:timestwo:maxlhs",
                "Too many output arguments.");
    }
    
    /* The input must be noncomplex doubles.*/
    mrows = mxGetM(prhs[0]);
    ncols = mxGetN(prhs[0]);
    if( !mxIsDouble(prhs[0]) || mxIsComplex(prhs[0]) ) {
        mexErrMsgIdAndTxt( "MATLAB:timestwo:inputNotRealScalarDouble",
                "Input must be noncomplex doubles.");
    }
    
    /* Assign pointer for the input matrix.*/
    x = mxGetPr(prhs[0]);
    lstar=x[0];
    //x[1]=floor(Nbins);
    Nbins=x[1];
    
    /* Create matrix for the return argument. */
    
    mrows_out=303; 
// row 0: average intensity for this angle bin; 
// row 1: average path length for this angle bin. 
// row 2: bin width for this angle bin. 
// rows 3-302: path length distribution for this angle bin.
    ncols_out=Nbins;  /* one column for each angle segment. The columns give a counter for each         */
    plhs[0] = mxCreateDoubleMatrix((mwSize)mrows_out, (mwSize)ncols_out, mxREAL);
    
    /* Assign pointer for the output matrix. */
    
    y = mxGetPr(plhs[0]);
    
    /* Call the pathlengths subroutine. */
    pathlengthsCyl(y,x,mrows_out,ncols_out);
}
\end{lstlisting}


\end{document}